\documentclass[letterpaper]{article} 
\usepackage{graphicx}
\usepackage{verbatim}
\usepackage{amssymb} 
\usepackage{enumitem}
\usepackage{tabularx}
\usepackage{natbib} 
\usepackage{color, colortbl}
\definecolor{Gray}{gray}{0.9}
\definecolor{pearl}{rgb}{0.94, 0.92, 0.84}
\definecolor{darkgreen}{rgb}{0.0, 0.2, 0.13}
\definecolor{amber}{rgb}{1.0, 0.49, 0.0}
\definecolor{or}{rgb}{0.99, 0.84, 0.69}
\usepackage{amsmath}
\usepackage{graphicx}
\usepackage[colorinlistoftodos]{todonotes}
\usepackage[colorlinks=true, allcolors=blue]{hyperref}
\usepackage{caption}
\usepackage[flushleft]{threeparttable}
\usepackage{siunitx}
\usepackage{breqn}
\usepackage{float}

\title{Environmental Monitoring Requirements for the ngVLA}
\author{T. K. Sridharan, Jeff Mangum \& Bryan Butler \\ NRAO}

\begin{document}
\pagestyle{plain} 
\pagenumbering{arabic}

\begin{figure}
\centering
\includegraphics[width=3.5cm]{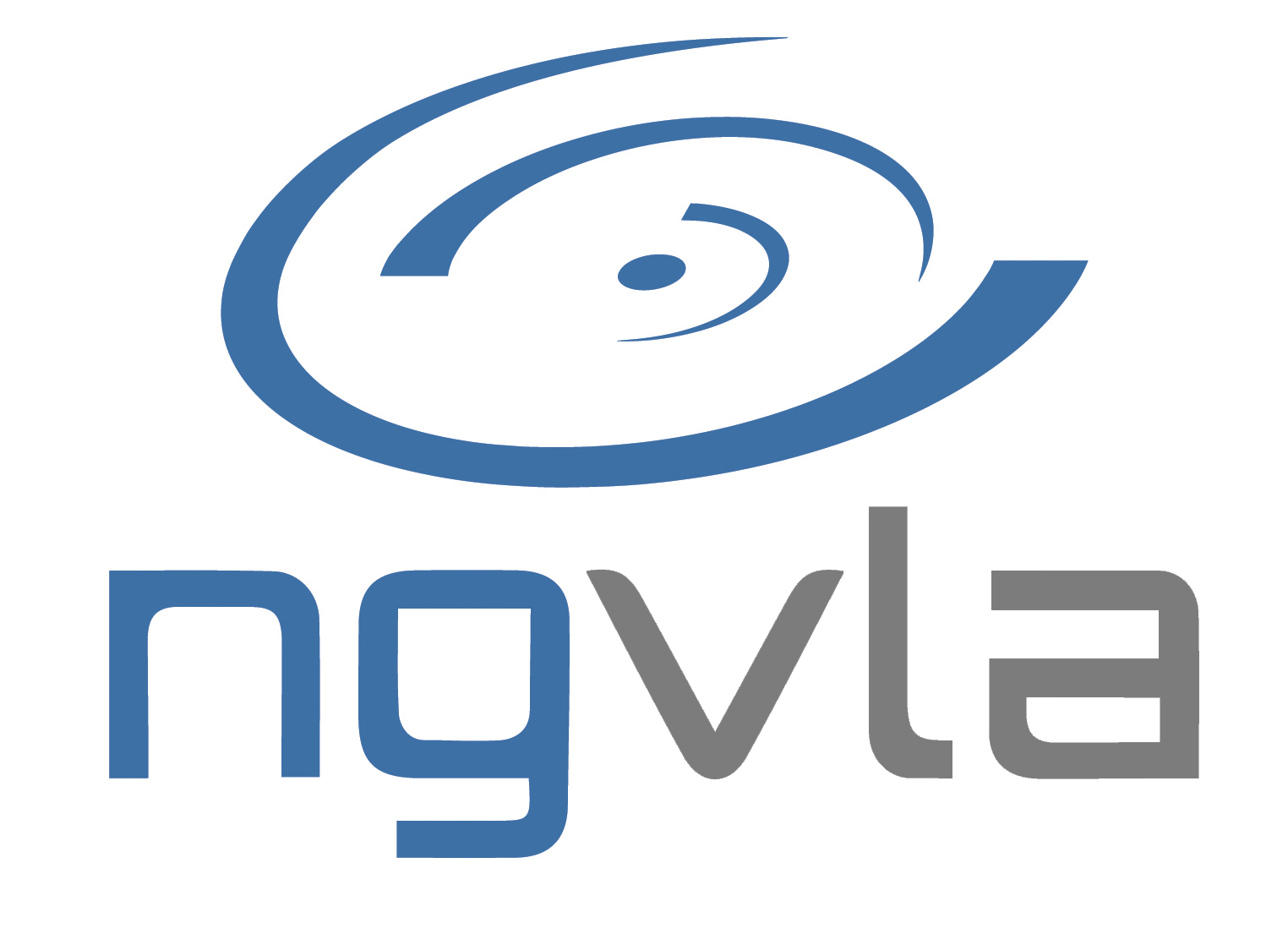}
{\center\Large\bf Memo No. 115 }
\end{figure}

\maketitle







\section{Introduction} 

Measurement of environmental parameters is one of the basic requirements for the proper operation of a telescope. Therefore, the  subject has been addressed by several observatories. This memo is intended to provide guidance for the measurement accuracy requirements in the context of the ngVLA. It relies on previous work for the EVLA \citep{ButlerPerley2008} and ALMA \citep{ALMAmemo366} and a review of the subject by \cite{Mangum2015PASP}. 

The local operational environment can be broadly divided into two categories: electromagnetic and physical. Meteorological parameters (weather) primarily constitute the physical environmental component and radio frequency interference (RFI) is the essential element of the electromagnetic environment. This memo focuses on the weather component and does not address the RFI, safety and physical infrastructure components.

Under weather, the relevant topics are (1) the correction to pointing arising from refraction in the atmosphere (2) the different delays in the arrival times of signals at different antennas due to propagation in the atmosphere (3) monitoring weather parameters to provide operations support, e.g. in determining prevalence of precision or normal conditions, dynamic scheduling and the choice of antennas to constitute a sub-array with a given set of characteristics, among others, and (4) archival. Here we restrict ourselves to the first two topics which impact the data obtained and its calibration.
\section{Pointing Sensitivity to Weather}



Closed form analytic expressions exist for the dependence of refraction correction to antenna pointing in elevation on surface temperature, pressure, and relative humidity at the location of the antenna (commonly used approximations of sufficient accuracy; reviews and references can be found in \cite{ALMAmemo366}, \cite{ButlerPerley2008} and \cite{Mangum2015PASP}, and are not repeated here). To estimate the sensitivity of the pointing correction to weather, we take the approach of deriving the partial derivatives of these expressions with respect to the respective weather parameters. 
Starting from \cite{Mangum2015PASP} (which also includes more information and citations to the origin of these equations), replacing $R_0$ with $\Delta z$ for the refraction correction (Equation 11) we have
\begin{equation}
\Delta z = 0.206265 N_0(ppm) \tan{z_o}~\textrm{(arcsec)}
\label{eq:deltazdef}
\end{equation}
and for refractivity (Equation 66):
\begin{eqnarray}
  ^{Rueger}N^{rad}_0 &=& 77.6890\frac{P_d}{T} + 71.2952\frac{P_w}{T} + 3.75463\times 10^5 \frac{P_w}{T^2}~\textrm{(ppm)} \nonumber \\
 &=& 77.6890\frac{P}{T} - 6.3938\frac{P_w}{T} + 3.75463\times 10^5 \frac{P_w}{T^2}~\textrm{(ppm)}
 \label{eq:norba}
 \end{eqnarray}
where
\begin{itemize}
  \item[$P_d$] is the partial pressure of dry gases in the atmosphere
(in hPa),
  \item[$P_w$] is the partial pressure of water vapor (in hPa),
  \item[$P$] is the total barometric pressure (in hPa), which is equal
    to $P_d + P_w$, and
  \item[$T$] is the ambient air temperature (in Kelvin).
  \item[$z_0$] is the zenith angle of the observation
  \item[$^{Rueger}N^{rad}_0$] is the radio refractivity (Rueger formulation) 
\end{itemize}

While the calculations for ngVLA requirements do not need the number of significant digits included here for the various coefficients, we retain full precision for completeness. Noting that the total atmospheric pressure $P$ must be measured by a barometer at the telescope site, $P_w$ may be calculated from the expression:
\begin{equation}
P_w = RH\frac{e_{sat}}{100}
\label{eq:pw}
\end{equation}
where $RH$ is the surface relative humidity (in percent) and $e_{sat}$ is the surface saturated water vapour pressure (in hPa).  There are two forms for $e_{sat}$ that we can use.  The first, from \cite{Crane1976} (which is used by \cite{ButlerPerley2008}), is given by:
\begin{eqnarray}
e^{crane}_{sat} &=& 6.105\exp\left[25.22\left(\frac{T-273}{T}\right) - 5.31\ln\left(\frac{T}{273}\right)\right]~\textrm{(hPa)} \nonumber \\
&=& 6.105\left(\frac{T}{273}\right)^{-5.31}\exp\left[25.22\left(\frac{T-273}{T}\right)\right]~\textrm{(hPa)}.
  \label{eq:rhesat-crane}
\end{eqnarray}
A second expression commonly used is given by \cite{Buck1981} (Equation C1 in \cite{Mangum2015PASP}):
\begin{equation}
e^{buck}_{sat} = \left(1.0007 + 3.46\times10^{-6}
  P\right)6.1121\exp\left[\frac{17.502\left(T-273.15\right)}{T-32.18}\right]~\textrm{(hPa)}.
  \label{eq:rhesat-buck}
\end{equation}
The difference between the \cite{Crane1976} and \cite{Buck1981} formalisms for $e_{sat}$ is $\lesssim 1$\% over the range of surface pressure and temperature from 500 to 1000\,mb and 255 to 300\,K, respectively, appropriate for the ngVLA.  In the following we will use the \cite{Buck1981} formalism for $e_{sat}$.

\begin{figure}
    \centering
    \includegraphics{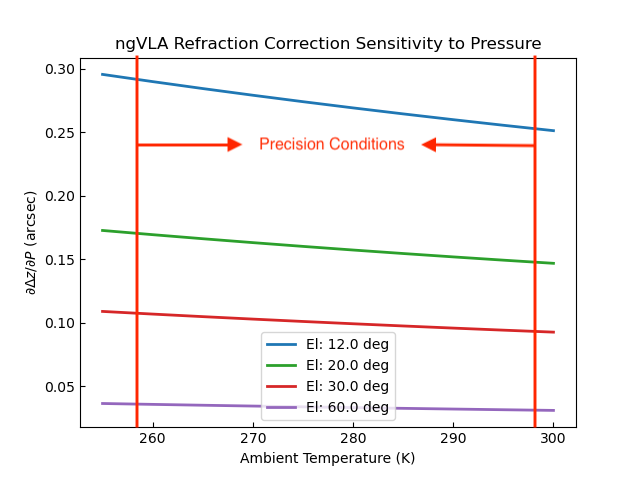}
    \caption{Refraction correction dependence on ground pressure (Equation~\ref{eq:partialP}) per mbar for a representative range of ground temperatures $T$ and observing elevations. The temperature range for precision conditions is marked. The required pointing correction accuracy of $<$0.5$^{\prime\prime}$ (Section~\ref{section:Req}) is outside the box and can be met with a 4.5 mbar pressure measurement accuracy for elevations $>$ 30$^{\circ}$ } 
    \label{fig:dzdp}
\end{figure}

We now derive the partial derivatives of Equation~\ref{eq:deltazdef} with respect to the respective weather parameters.  We start with the partial derivative with respect to total pressure $P$.  Noting that $P_w = P - P_d$ and that $P_d \gg P_w$, so that $\frac{\partial P_d}{\partial P} \sim 1$ and $\frac{\partial P_w}{\partial P} \ll 1$, only the first term in Equation~\ref{eq:norba} is retained:
\begin{eqnarray}
\frac{\partial{\Delta z}}{\partial{P}} &=& 0.206265 \tan{z_o} \left(\frac{77.6890}{T}\right)~\textrm{(arcsec)} \nonumber \\
&\simeq& \left(\frac{16.024}{T}\right)\tan{z_o}~\textrm{(arcsec)}
\label{eq:partialP}
\end{eqnarray}
where we have ignored the insignificant dependence of $e_{sat}$ on the pressure $P$.  Figure~\ref{fig:dzdp} shows Equation~\ref{eq:partialP} for a representative range of ground temperatures.

\begin{figure}
    \centering
    \includegraphics{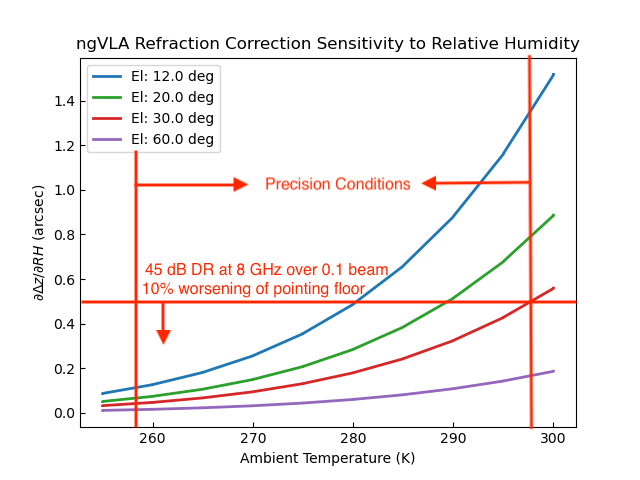}
    \caption{Refraction correction dependence on ground relative humidity (Equation~\ref{eq:partialRH}), per 1\% for a representative range of ground temperatures $T$, pressures $P$ (whose effect is not discernible) and observing elevations. The temperature range for precision conditions is marked. The required pointing correction accuracy of $<$0.5$^{\prime\prime}$ shown (section~\ref{section:Req}) can be met with a 1\% relative humidity measurement accuracy for elevations $>$ 30$^{\circ}$ }
    \label{fig:dzdrh}
\end{figure}

The partial derivative with respect to relative humidity $RH$ (in percent) requires substitution of Equation~\ref{eq:pw} into Equation~\ref{eq:norba}:
\begin{eqnarray}
\frac{\partial{\Delta z}}{\partial{RH}} &=& 0.206265 \tan{z_o} \left(\frac{77.6890 e^{buck}_{sat}}{100 T} - \frac{6.3938 e^{buck}_{sat}}{100 T} + \frac{3.75463\times 10^5 e^{buck}_{sat}}{100 T^2}\right)~\textrm{(arcsec)} \nonumber \\
&\simeq& \tan{z_o}\left(\frac{0.160 e^{buck}_{sat}}{T} -\frac{0.013 e^{buck}_{sat}}{T} + \frac{774.45 e^{buck}_{sat}}{T^2}\right)~\textrm{(arcsec)} \nonumber \\
&\simeq& \tan{z_o}\left(\frac{0.147 e^{buck}_{sat}}{T} + \frac{774.45 e^{buck}_{sat}}{T^2}\right)~\textrm{(arcsec)}
\label{eq:partialRH}
\end{eqnarray}
with $e^{buck}_{sat}$ given by Equation~\ref{eq:rhesat-buck}. There is a negligibly weak dependence of the partial derivative on pressure through $e^{buck}_{sat}$. Figure~\ref{fig:dzdrh} shows Equation~\ref{eq:partialRH} for a representative range of ground temperatures and pressures.

\begin{figure}
    \centering
    \includegraphics{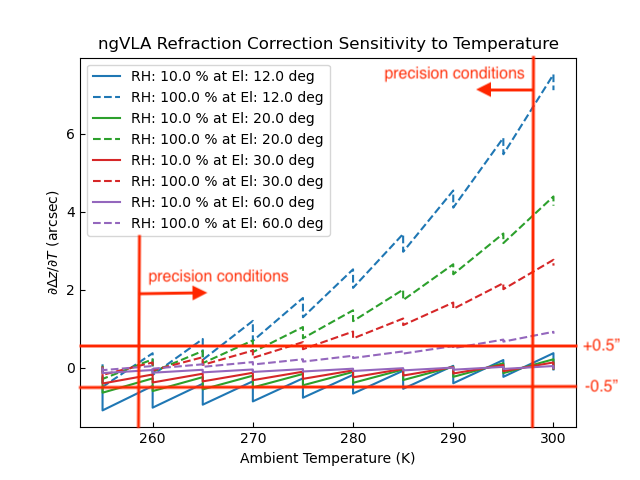}
    \caption{Refraction correction dependence on ground temperature (Equation~\ref{eq:partialTfinal}) for a representative range of ground temperatures $T$, pressure $P$, observing elevations and two relative humidities. In order to present the $T$ and $P$ dependences in a single plot, 5 deg intervals in $T$ are each spanned by a 500-1100 hPa range in pressure leading to the saw tooth patterns.   The temperature range for precision conditions is marked. The required pointing correction accuracy of $<$0.5$^{\prime\prime}$ shown (Section~\ref{section:Req}) is readily met at low humidities.  A 0.3$^{\circ}$ temperature measurement accuracy meets the requirements for elevations $>$ 30$^{\circ}$, even at 100\% humidity.}
    \label{fig:dzdt}
\end{figure}

Finally, the partial derivative with respect to temperature $T$ (in Kelvin) is a bit more complicated to derive as both $N_0$ and $P_w$ (through the temperature dependence in $e^{buck}_{sat}$) have dependencies on $T$:
\begin{eqnarray}
\frac{\partial{\Delta z}}{\partial{T}} &=& 0.206265 \tan{z_o} \left[\frac{\partial}{\partial{T}}\left(77.6890\frac{P_d}{T}\right) + \frac{\partial}{\partial{T}}\left(71.2952\frac{P_w}{T}\right) + \frac{\partial}{\partial{T}}\left(3.75463\times 10^5 \frac{P_w}{T^2}\right)\right] \nonumber \\
&\simeq& \tan{z_o} \left[\frac{\partial}{\partial{T}}\left(16.02\frac{P_d}{T}\right) + \frac{\partial}{\partial{T}}\left(14.71\frac{P_w}{T}\right) + \frac{\partial}{\partial{T}}\left(7.74\times 10^4 \frac{P_w}{T^2}\right)\right]~\textrm{(arcsec)}
\label{eq:partialT}
\end{eqnarray}
The derivative of the second term in Equation~\ref{eq:partialT} is given by:
\begin{eqnarray}
\frac{\partial}{\partial{T}}\left(14.71\frac{P_w}{T}\right) &=& 14.71\left(\frac{T\frac{\partial}{\partial{T}}\left(RH\frac{e^{buck}_{sat}}{100}\right) - RH\frac{e^{buck}_{sat}}{100}}{T^2}\right) \nonumber \\
&=& 0.1471 RH \left(\frac{T\frac{\partial{e^{buck}_{sat}}}{\partial{T}} - e^{buck}_{sat}}{T^2}\right)
\label{eq:partialTterm2}
\end{eqnarray}
with the partial derivative of $e_{sat}$ with respect to $T$ given by:
\begin{eqnarray}
\frac{\partial{e^{buck}_{sat}}}{\partial{T}} &=& \left(1.0007 + 3.46\times10^{-6}
  P\right)6.1121\exp\left(\frac{17.502\left(T-273.15\right)}{T-32.18}\right)\left(\frac{4217.46}{(T - 32.18)^2}\right) \nonumber \\
  &=& \frac{4217.46}{(T - 32.18)^2} e^{buck}_{sat}~\textrm{(hPa)}.
  \label{eq:partialesat}
\end{eqnarray}
The derivative of the third term in Equation~\ref{eq:partialT} is given by:
\begin{eqnarray}
\frac{\partial}{\partial{T}}\left(7.74\times10^4\frac{P_w}{T^2}\right) &=& 7.74\times10^4\left(\frac{T^2\frac{\partial}{\partial{T}}\left(RH\frac{e^{buck}_{sat}}{100}\right) - 2TRH\frac{e^{buck}_{sat}}{100}}{T^4}\right) \nonumber \\
&=& 774 RH \left(\frac{T\frac{\partial{e^{buck}_{sat}}}{\partial{T}} - 2e^{buck}_{sat}}{T^3}\right)
\label{eq:partialTterm3}
\end{eqnarray}
with the partial derivative of $e_{sat}$ with respect to $T$ given by Equation~\ref{eq:partialesat}.  Combining Equations~\ref{eq:partialTterm2} and \ref{eq:partialTterm3} into Equation~\ref{eq:partialT} we find:
\begin{eqnarray}
\frac{\partial{\Delta z}}{\partial{T}} &\simeq& \tan{z_o} \left[-16.02\frac{P_d}{T^2} + 0.1471 RH \left(\frac{T\frac{\partial{e^{buck}_{sat}}}{\partial{T}} - e^{buck}_{sat}}{T^2}\right) + 774 RH \left(\frac{T\frac{\partial{e^{buck}_{sat}}}{\partial{T}} - 2e^{buck}_{sat}}{T^3}\right)\right] \nonumber \\
&\simeq& \tan{z_o} \left[\frac{-16.02}{T^2}\left(P - P_w\right) + 0.1471 RH \left(\frac{T\frac{\partial{e^{buck}_{sat}}}{\partial{T}} - e^{buck}_{sat}}{T^2}\right) + 774 RH \left(\frac{T\frac{\partial{e^{buck}_{sat}}}{\partial{T}} - 2e_{sat}}{T^3}\right)\right] \nonumber \\
&\simeq& \frac{-16.02\tan{z_o}}{T^2} \left[\left(P - P_w\right) - 0.009 RH \left(T\frac{\partial{e^{buck}_{sat}}}{\partial{T}} - e^{buck}_{sat}\right)
 - 48.31 RH \left(\frac{T\frac{\partial{e^{buck}_{sat}}}{\partial{T}} - 2e^{buck}_{sat}}{T}\right)\right]~\textrm{(arcsec)} \nonumber \\
\label{eq:partialTfinal}
\end{eqnarray}
with $P_w$ and the derivative of $e^{buck}_{sat}$ with respect to $T$ given by Equations~\ref{eq:pw} and \ref{eq:partialesat}, respectively.  Figure~\ref{fig:dzdt} shows Equation~\ref{eq:partialTfinal} for a representative range of ground temperatures and pressures.



As a check of this derivation we have verified numerical agreement between these expressions and the computational results for specific cases presented in \cite{ALMAmemo366} for ALMA, where we find agreement to $< 0.03^{\prime\prime}$. We also find them to be consistent with the results of a numerical variational analysis presented in \cite{ButlerPerley2008} for the EVLA. 

These expressions are used below to determine the accuracies to which the weather parameters need to be measured. As can be seen, the sensitivities for humidity and pressure only depend on temperature and the temperature sensitivity depends on all  three weather parameters. The functional dependence of all sensitivities on elevation is $tan(z)$. 

\section{Refraction Correction Error Requirement}\label{section:Req}

The pointing specification and expected performance of the prototype antenna are taken as indicative references for feasibility. The 3$^{\prime\prime}$ pointing specification for the prototype antenna under precision conditions (\cite{AntTechReq22}) translates to a primary beam gain factor as high as 0.97 even at the highest operating frequencies. Given the stringent ngVLA image dynamic range requirements (\cite{SysReq20}), the main consideration, therefore, is the impact of inadequate refraction correction on the dynamic range achieved in the constructed images.  
 

We adopt two guiding principles to assess the allowable error in the refraction correction to elevation pointing calculated from weather parameters: (1) the error should not limit dynamic range to below 45 dB at 8 GHz in the central region of the primary beam without requiring pointing self calibration. This is estimated to be 0.5$^{\prime\prime}$ within 10\% of the primary beam half-power beamwidth for 107 antennas (achieving 45 dB over the full HPBW requires pointing self-calibration; \cite{Jagan23}). (2) the error should not limit the best pointing performance achieved by the antennas under windless conditions. Numerically, requiring that the increase in pointing error due to this correction be less than 10$\%$ of the best pointing errors under windless conditions that the antenna control systems are able to deliver, which is 1.2$^{\prime\prime}$ for the prototype antenna, also results in a 0.5$^{\prime\prime}$ limit. We adopt 0.5$^{\prime\prime}$ as the pointing error correction accuracy requirement for further discussions.



The assessment here only considers the precision operating conditions temperature range of 258-298K, relevant for the most stringent ngVLA performance requirements. Next, although not explicitly identified as a science or system requirement, we consider the fact that the highest dynamic range performance will not be required over the full elevation range, in particular at elevations as low as 12 deg. Specifically, for the dynamic range requirement which drives pointing requirement we adopt an elevation limit of 30 deg. 

Figures \ref{fig:dzdp}, \ref{fig:dzdrh} and \ref{fig:dzdt} already presented the adopted 0.5$^{\prime\prime}$ requirement and the precision condition temperature range. As can be seen from these figures and the above discussion, measurement accuracies of 4.5 mbar in pressure, 1\% in relative humidity and 0.3 deg in temperature meet the adopted requirements and are therefore identified as the required surface weather parameter measurement accuracies.   

 We note that the ngVLA science and system  requirements do not specify an elevation limit for the achievement of the image dynamic range requirements. Additionally, both reference pointing and pointing self-calibration, which is the strategy adopted for the attainment of the image dynamic range requirements, will also correct for residual errors in refraction correction \citep{Jagan23, Bhatnagar2017AJ}. The requirements presented here will eliminate refraction correction as a limiting factor in cases where pointing self-calibration may not be viable. 


Based on the above  analysis, the  weather parameter measurement accuracies summarized below are recommended, where an elevation limit $>$ 30\,deg is adopted: \\
 
\noindent 
\begin{tabular}{l r l}
(1) Temperature: & $\Delta T$ = & 0.3 K  \\
(2) Relative humidity: & $\Delta RH$ = & 1\%  \\
(3) Pressure: & $\Delta P$ = & 4.5 mbar  \\
\end{tabular} 
\\

No recommendations for wind measurement requirements, which are expected to be coarse, are provided, as they pertain primarily to operational aspects of safety and the determination of the prevalence of precision vs normal conditions.

\section{Weather Station Distribution}

The distribution of weather stations needed to provide the required measurements for all antennas in the array depends on the lateral distance scale over which the parameters change by the amounts derived above as the measurement accuracy requirements. In general, in a well mixed, isothermal atmosphere we do not expect significant variations of weather parameters from antenna to antenna in the inner core area. Altitude difference would be expected to be the primary cause of variations in addition to location specific atmospheric boundary layer and geographical effects. While future ground truth measurements at the chosen sites should guide the distribution of weather stations as we are not aware of available measurements with sufficient spatial granularity, we can arrive at some broad conclusions using general considerations. 

Using a tropospheric adiabatic lapse rate of 9.8 K/km for dry air (worst case; lower for wet air), a 0.3 K change occurs with an altitude change of 30 m. Similarly, a pressure change of 5 mbar (or hPa) occurs with an altitude variation of 55 m  using a scale height of 8.5 km for the earth's atmosphere at a pressure of 780 mbar applicable to the ngVLA sites. The temperature variation provides a more stringent altitude variation range of $\pm$30m. We recommend  that antennas within a $\pm$30 m ($\pm$100 ft) range in altitude be grouped to share a weather station. As already noted, while ground truth measurements should ultimately determine the distribution, we expect that the 114 antennas of the Main Array within a 2.2 km radius and the SBA can be served by a small number (perhaps one) of weather stations. Each of the Long Baseline Array antennas will clearly need separate stations. The choice for the remaining Main Array antennas (spiral) will critically depend on future measurements to characterize variations, with suitably grouped antennas sharing a set of weather stations. Interpolation of weather data from the distribution of stations can provide additional granularity for application to individual antennas.   

\section{Delay Correction Error Requirements}

We now turn to the measurement accuracy of surface weather parameters arising from delay estimation requirements.  
Bulk delays which will be applied to data from each antenna before correlation 
include an important contribution from the atmosphere. This is typically estimated from measured 
surface meteorological parameters (temperature, dew point temperature 
or humidity, and pressure) and an assumed atmospheric vertical structure 
above the antenna.  This structure usually incorporates hydrostatic 
equilibrium for the bulk atmosphere with some assumed scale height
and different scale heights for O$_2$ and H$_2$O (water vapor), which are 
the primary atmospheric constituents of relevance at the wavelengths of interest here (see references in \cite{ButlerPerley2008} and \cite{Mangum2015PASP}.

Past investigations of the measurement accuracy needed for the 
weather parameters to be used as inputs to the atmospheric delay 
calculation for the VLBA and VLA have treated this in various ways 
\citep{Clark1986,Clark1987,ButlerPerley2008} but all have come to a 
similar conclusion - the driving consideration for the measurement accuracy of 
surface meteorological parameters is in the refraction (pointing)
correction.  The main point is that once you get to delay accuracy 
levels that drive the surface quantity measurement accuracy to better 
than the refraction calculation, uncertainties in the vertical 
structure of the atmosphere (mostly the water vapor), which are not captured by the surface measurements, dominate the 
uncertainty in the overall delay calculation, rather than the measurement accuracies 
of the surface parameters.  We recognize these previous conclusions and do not perform a detailed analysis of 
uncertainties and requirements from a delay estimate perspective. While the measured 
surface weather parameters should still be used in the bulk delay calculation, the 
accuracy of their measurements is not driven by this consideration.  
Whatever is required for the refraction calculation is good enough for 
the purposes of estimating the atmospheric delay.  
 
We note that the surface weather parameters alone will not be used to 
determine delays and observations of delay calibrators and further corrections from periodic observations of complex gain (amplitude and phase vs. time) 
calibrators will be sufficient for calibrating delay (or phase).  Where self-calibration can 
be used, even those are not necessarily needed, although observations will
certainly include calibrator measurements.

\section{Measurement Cadence}

The fastest measurement cadence requirement arises from the expected use of surface weather data as empirical inputs to water vapor radiometry. While the details of the tropospheric delay correction algorithm are yet to be worked out, we anticipate seconds time scale as appropriate from this perspective. Accordingly, we set 1 second as the measurement cadence requirement.

\section{Weather Sensor Location}

\noindent From the pointing correction perspective the conditions that matter are the ones closest the antenna including any influence from the antenna itself. Unlike delay, the amount (angle) of refraction correction only depends on the refractive index at the final wavefront location and is independent of the detailed structure of the refractive index distribution along the line of sight (\cite{Smart1962}, \cite{Mangum2015PASP} and references therein). The sensors can be mounted either on the antenna or on a nearby tower such that the measurements correspond to the mean height of the aperture/focal plane. We leave this decision to implementation convenience. We note, however, that wind sensors mounted on the antenna may suffer from shadowing effects and the modification of the flow due to the antenna structure which will have to be considered before a final decision. This may be evaluated through a comparison of data from instruments mounted on the prototype antenna and on a nearby tower.

\section{Recommended Requirements}

The recommended weather parameter measurement requirements are summarized below:\\
 
\noindent 
\begin{tabular}{l r l}
(1) Temperature: & $\Delta T$ = & 0.3 K  \\
(2) Relative humidity: & $\Delta RH$ = & 1\%  \\
(3) Pressure: & $\Delta P$ = & 4.5 mbar  \\
(4) Measurement cadence: & & 1 sec \\
(5) Station distribution: & & TBD using ground truth measurements \\
(6) Sensor location: & & TBD using prototype antenna tests \& implementation convenience
\end{tabular} 
\\

\bibliographystyle{aasjournal}
\bibliography{ngVLAEnvironMonitoring}{}

\end{document}